\documentclass[aps,prl,twocolumn,superscriptaddress]{revtex4-1}

\usepackage{latexsym}
\usepackage{amsmath, amsthm, amssymb}
\usepackage{mathrsfs}
\usepackage{epsfig}
\usepackage{graphicx}
\usepackage{dcolumn}
\usepackage{expl3}
\usepackage{float}

\usepackage{tikz}
\usetikzlibrary{trees}

\usepackage{natbib}
\usepackage{hyperref}
\usepackage{placeins}

\allowdisplaybreaks

\begin{document}
\title{Contact-mediated cellular communication
supplements positional information to regulate spatial patterning during development}

\author{Chandrashekar Kuyyamudi}
\affiliation{The Institute of Mathematical Sciences, CIT Campus, Taramani, Chennai 600113, India}
\affiliation{Homi Bhabha National Institute, Anushaktinagar, Mumbai 400 094, India}
\author{Shakti N. Menon}
\affiliation{The Institute of Mathematical Sciences, CIT Campus, Taramani, Chennai 600113, India}
\author{Sitabhra Sinha}
\affiliation{The Institute of Mathematical Sciences, CIT Campus, Taramani, Chennai 600113, India}
\affiliation{Homi Bhabha National Institute, Anushaktinagar, Mumbai 400 094, India}
\date{\today}

\begin{abstract}
Development in multi-cellular organisms is marked by a high degree of spatial organization of the cells attaining distinct fates in the embryo. We show that receptor-ligand interaction between cells in close physical proximity adaptively regulates the local process of selective gene expression in the presence of a global field set up by a diffusing morphogen that provides positional cues. This allows information from the cellular neighborhood to be incorporated into the emergent thresholds of morphogen concentration that dictate cell fate, consistent with recent experiments.
\end{abstract}
\maketitle

Spatial symmetry breaking is a fundamental prerequisite to morphogenesis, or the development of
form, in living organisms, such that an initially homogeneous domain exhibits patterns
in the concentrations of molecular species referred to as morphogens~\cite{Cross1993,Koch1994,Ball1999,Zhang2018}.
This can come about through either self-organizing reaction-diffusion processes~\cite{Turing1952,Meinhardt1982,Werner2015}
or from the anisotropy associated with the concentration gradient of a morphogen produced
by a localized source~\cite{Wolpert1969,Wolpert1989,Sharpe2019}.
While in the simplest scenario involving the latter mechanism,
the morphogen diffuses through space subject to
uniform linear degradation~\cite{Crick1970,Driever1988,Teleman2001,Lander2002,Lander2007}, more complex
means of establishing a morphogen gradient have been proposed~\cite{Bollenbach2005,England2005,Hornung2005,Ben2010,Yuste2010,Muratov2011}.
Cells attain different fates according to the positional information provided by the
local concentration of the morphogen
vis-a-vis threshold values that emerge from the dynamics of the interpretation
module of their genetic regulatory network~\cite{Gurdon2001,Ashe2006,Rogers2011,Gilbert2013}.
However, the spatial pattern of cell fates is not entirely determined by these local interactions
as recent experiments have highlighted the role of cell-cell communication in this process~\cite{Kong2015}.

Cells in the developing embryo are known to interact with other cells that are in
close physical proximity through contact-mediated signaling.
This can occur through binding between membrane-bound receptors and ligands
on the surfaces of neighboring cells, a prominent example being the
evolutionarily conserved Notch signaling
pathway~\cite{Artavanis1999}. Notch-mediated interactions, that are
believed to have arisen early in evolution, have been shown to play a
crucial role in the development of all metazoans~\cite{Artavanis1999,Kopan2009}.
It has been demonstrated to help sharpen the boundaries between
regions having different cell fates in the presence of
fluctuating morphogen concentrations~\cite{Sprinzak2011}, providing an important mechanism
for systems to be robust with respect to noisy signals~\cite{Erdmann2009,Lander2011,Lander2013}.
More importantly, Notch signaling is capable of self-regulation as the signaling between
neighboring cells implements an effective feedback loop~\cite{note1}.

\begin{figure}[htbp!]
\includegraphics[width=0.5\textwidth]{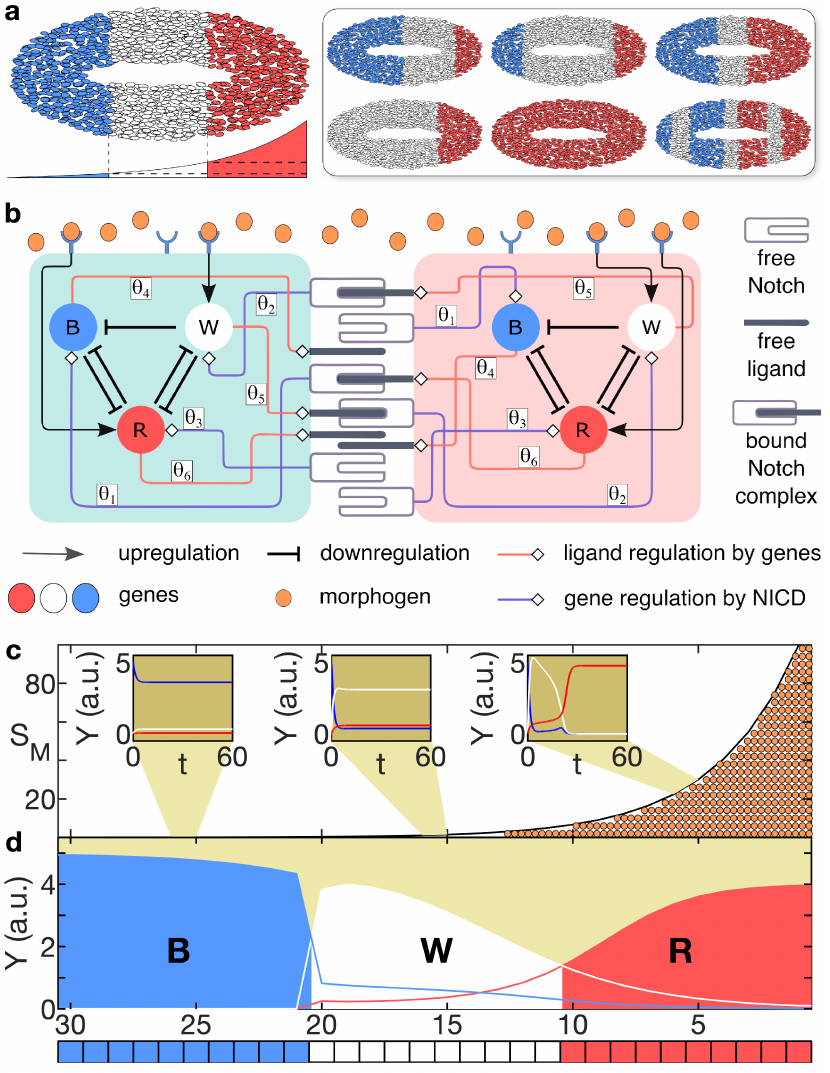}
\caption{\textbf{Contact-mediated signaling regulates the differential expression of cell fates dictated by morphogen concentration profiles.}
(a)~Schematic diagrams illustrating the French Flag problem, namely, how positional information provided by spatial gradients of
morphogen concentration specify patterns of cell fates in embryonic tissue.
Equally sized domains of cells exhibiting one of three different fates, viz., blue, white and red, characterize the idealized situation (left),
shown for the case of patterning in the vertebrate neural tube by a gradient of Sonic hedgehog morphogen
(whose concentration profile is displayed).
Under different conditions, variations preserving the chromatic order and number of fate boundaries of the
idealized situation can arise (a: right, top row); however, other variations may violate these (a: right, bottom row).
(b)~Schematic diagram of a pair of cells coupled via Notch signaling in the presence of an external morphogen.
Each cell contains a morphogen interpretation module comprising a regulatory circuit of fate-inducing genes B, W and R.
Notch intra-cellular domains (NICD), released upon successful binding of Notch receptors to ligands from the neighboring cell,
affect expression of B, W and R with strengths $\theta_{1,2,3}$, respectively. This in turn regulates the
production of Notch ligand with strengths $\theta_{4,5,6}$.
(c)~Spatial variation of the response $S_M$ to the morphogen across a one-dimensional domain comprising $30$ cells.
The three insets display the time evolution
of gene expression levels Y ($=$ B, W or R, in arbitrary units) for cells that are
subject to low, intermediate and high morphogen concentrations,
respectively.
(d) The resulting final expression levels Y of the patterning genes.
The maximally expressed gene at each cell determines its fate, as
shown in the schematic representation of the $1$D domain displayed at the bottom.}
\label{figure1}
\end{figure}

In this paper we present a plausible mechanistic basis for explaining how inter-cellular interactions influence cell fate determination, as indicated by recent experiments, e.g.,
on the mouse ventral spinal cord~\cite{Kong2015}, by allowing Notch to alter the
expression of genes in the morphogen interpretation module, which in turn control
the production of Notch ligands.
Using a three-gene interpretation module associated with the Sonic
Hedgehog (Shh) morphogen gradient in vertebrate neural
tubes~\cite{Dessaud2008,Dessaud2010,Balaskas2012}, we show that
specific types of Notch-mediated coupling allow the size of the domains
corresponding to different cell fates
to be varied in a regulated manner. They retain the broad features of the
reference pattern obtained
in the absence of any intercellular coupling, while avoiding phenotypes
that do not preserve the number and the sequence of these domains [Fig.~\ref{figure1}~(a)].
Our results suggest that the emergent thresholds for the
morphogen concentration that determine the localization of various cell fates are
not only an outcome of the interaction between the morphogen gradient
manifested across an entire embryonic segment with the gene circuit dynamics at the cellular
level, but also the
intermediate-scale dynamics of inter-cellular interactions.

To investigate how the spatial patterning of cell fates are affected by juxtacrine signaling, we consider a linear array of cells responding to a morphogen whose concentration decays exponentially away from the source~\cite{Crick1970,note2}.
This spatial profile is reflected in the response of the cells in terms of the concentration
of the downstream signaling molecules released as a result of binding of morphogen molecules to
receptors on the cell membrane, viz., $S_M (x) = S_M (0) {\rm exp} (-x/\lambda_M)$, where $x$ is the distance of a cell
from the source of the morphogen.
The external signal concentration sensed by each cell through its receptors affects the expression of a set of genes that functions
as the morphogen interpretation module.
We choose one that has been proposed in the context of
vertebrate neural tube patterning, comprising the genes Pax6, Olig2 and Nkx2.2, in the presence of a Sonic hedgehog (Shh)
morphogen gradient~\cite{Balaskas2012}.
Fig.~\ref{figure1}~(b) shows the module with the regulatory motif of three patterning genes B, W and R,
that mutually repress each other, with the sole exception of W by B. The gene having the highest expression level in each cell
determines its fate, indicated by blue, white or red, which correspond to genes B, W and R, respectively [Fig.~\ref{figure1}~(c-d)].
As Pax6 is the only gene whose expression occurs even in the absence of the Shh morphogen, we consider this pre-patterning gene (B)
to be expressed at very high levels initially, in contrast to the other two.
The time-evolution of the expression of the three genes are described by:
\begin{align}
  \label{eq1}
  \frac{dB}{dt} &=
  \frac{\alpha + \varphi_{1}\left.\frac{N^b}{K_N}\right.}{1 +
  \left(\frac{R}{K}\right)^{h_1} + \left(\frac{W}{K}\right)^{h_2} +
  \xi_{1}\left.\frac{N^b}{K_N}\right.} - k_{1}B\,,\\
   \label{eq2}
   \frac{dW}{dt} &=
   \frac{\beta S_M + \varphi_{2} \left.\frac{N^b}{K_N}\right.}{1 + S_M +
   \xi_{2}\left.\frac{N^b}{K_N}\right.}~
   \frac{1}{1 + \left(\frac{R}{K}\right)^{h_3}} - k_{2}W\,,\\
   \label{eq3}
   \frac{dR}{dt} &=
   \frac{\gamma S_M + \varphi_{3} \left.\frac{N^b}{K_N}\right.}{1 + S_M +
   \xi_{3}\left.\frac{N^b}{K_N}\right.}~
   \frac{1}{1 + \left(\frac{B}{K}\right)^{h_4} +
   \left(\frac{W}{K}\right)^{h_5}} - k_{3}{R}\,,
\end{align}
where $\alpha, \beta , \gamma$ are the maximum growth rates and $k_{1,2,3}$ are the decay rates
of expression for the three genes, while $K$, $K_N$ and $h_{1,2,3,4,5}$ specify the nature of the response functions.
The parameters $\varphi_{1,2,3}$ and $\xi_{1,2,3}$ are associated with the juxtacrine coupling of adjacent cells through
the canonical Notch signalling pathway~\cite{Artavanis1999,Kopan2009}. To describe the dynamics resulting from the coupling, Eqs.~(\ref{eq1})-(\ref{eq3}) are augmented with the time-evolution equations of the concentrations $L$ and $N^b$ of the
Notch ligand and the Notch intra-cellular domain (NICD), respectively:
\begin{align}
   \frac{dL}{dt} &=
   \frac{\beta_L + \phi_{4}\frac{B}{K} + \phi_{5}\frac{W}{K} + \phi_{6}\frac{R}{K}}{1
   + \zeta_{4}\frac{B}{K} + \zeta_{5}\frac{W}{K} + \zeta_{6}\frac{R}{K}}
   - \frac{L}{\tau_L}\,,\\
   \frac{dN^b}{dt} &=
   \frac{\beta_{N^b}L^{trans}}{K + L^{trans}} - \frac{N^b}{\tau_{N^b}}\,.
\end{align}
Here, the parameters $\beta_{L,N^b}$ and $\tau_{L,N^b}$ correspond to the maximum growth rates and mean lifetimes for the ligand
and NICD, respectively.
The binding of Notch receptors of a cell to corresponding ligands of neighboring cells ($L^{trans}$) causes the receptor's intracellular domain
to be released and translocated to the nucleus~\cite{note3}.
We consider Notch and the patterning genes to regulate the expression of each other [see Fig.~\ref{figure1}~(b)].
Specifically, we consider four classes of inter-cellular interactions based on whether NICD up or downregulates the expression of B, W and R genes,
and whether Notch ligand production is promoted or repressed by the patterning genes
(mirroring the response of the ligands Jagged and Delta, respectively~\cite{Shimojo2011,Manderfield2012,Boareto2015}).
For simplicity, the ligand is assumed to be either activated by
all the genes or inhibited by each of them, while the genes themselves are regulated by NICD in
a qualitatively identical manner.
Thus, the four classes of inter-cellular coupling, defined by up ($G^+$) or downregulation ($G^-$) of the patterning genes,
and promotion ($L^+$) or repression ($L^-$) of the ligand, and specified by the parameter set $(\varphi_{i},\xi_{i},\phi_{j},\zeta_{j})$,
correspond to type I: $G^-$, $L^-$
 $(0,\theta_{i},0,\theta_{j})$; type II: $G^-$, $L^+$ $(0,\theta_{i},\theta_{j},1)$; type III: $G^+$, $L^-$
$(\theta_{i},1,0,\theta_{j})$; and type IV: $G^+$, $L^+$ $(\theta_{i},1,\theta_{j},1)$, where $i=1,2,3$ and $j=4,5,6$.

We choose values for the parameters such that the absence of coupling (i.e., $\varphi_i =0$, $\xi_i= 0$, $\forall i$)
yields an idealized flag with three chromatic regions of equal length, each corresponding to distinct cell fates (see SI).
To see how Notch signalling between adjacent cells
can alter the ordered pattern
of cells having different fates,
even when the morphogen gradient and the parameters of the interpretation module are kept unchanged,
we systematically investigate the six-dimensional parameter space spanned
by $\Theta =\, \{\theta_1,\theta_2,\theta_3,\theta_4,\theta_5,\theta_6\}$.
For each of the four types of coupling described above,
we consider $10^5$ realizations of the model obtained by randomly sampling $\Theta$.
Each of the parameters $\theta_{1,\ldots,6}$
is sampled from the interval $[1,10]$ (for $G^+$ and $L^+$) or $[0.1,1.0]$ (for $G^-$ and $L^-$).
\begin{figure}[htbp!]
\includegraphics[width=0.5\textwidth]{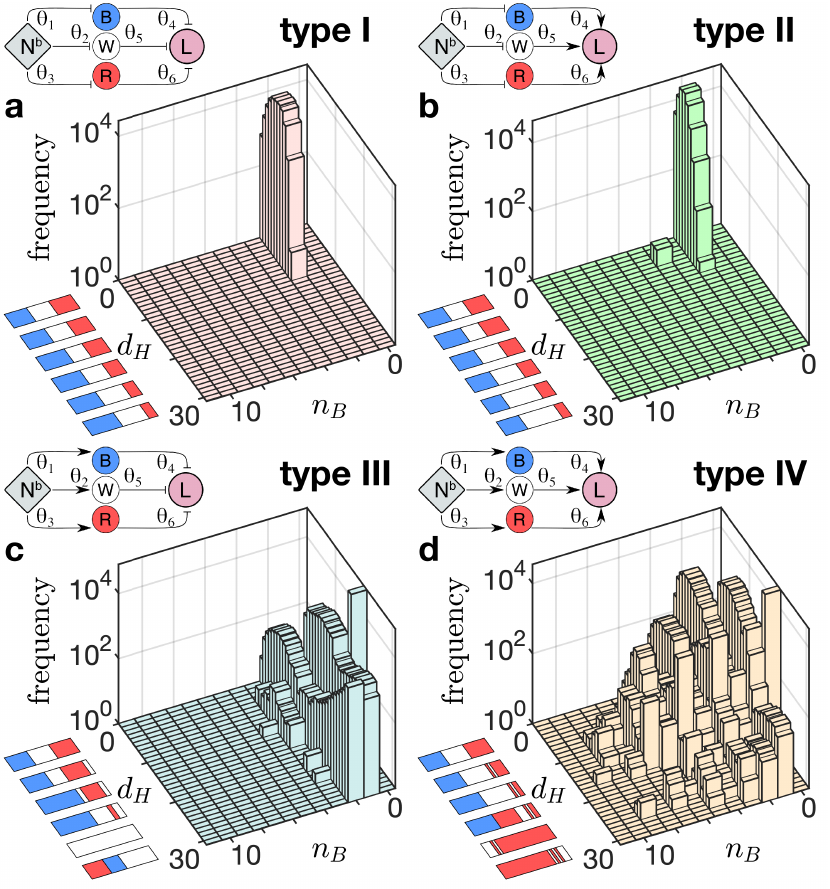}
\caption{\textbf{The diversity in the spatial patterns of cell fates is controlled by the nature of interactions underlying Notch-mediated inter-cellular coupling.}
The inter-cellular
interactions can be classified into four types, determined by whether NICD up or
downregulates the patterning genes, and in turn, the genes
up or downregulate ligand production, represented by the four motifs in the upper left corners in (a)-(d) [arrows representing
up/downregulation are as indicated in Fig.~\ref{figure1}~(b)].
For each type, the frequency distributions of different flags, i.e., patterns representing the sequential arrangement of distinct cell fates, are obtained by
randomly sampling $\theta_{1,\ldots,6}$, are shown in (a)-(d) for a $1$D domain comprising $30$ cells subject to the morphogen
gradient shown in Fig.~\ref{figure1}~(c).
In the absence of inter-cellular coupling, the domain is divided into three equal segments
of cells having different fates
[as in Fig.~\ref{figure1}~(d)].
The flags obtained upon coupling the cells are characterized by the number of fate boundaries $n_B$ and the difference $d_H$
with the pattern in the uncoupled system (which has equal chromatic divisions).
For types I, II (where NICD downregulates B, W and R) almost all flags have the same
chromatic order and $n_B$ as the idealized flag shown in Fig.~\ref{figure1}~(a, left), with $d_H$ limited to very low values
[a and b, cf. Fig.~\ref{figure1}~(a, right, top row)].
In contrast, the flags seen for types III, IV  (where NICD upregulates B, W
and R) exhibit large variation from the uncoupled case in terms of both $d_H$ and
$n_B$ [c and d, cf. Fig.~\ref{figure1}~(a, right, bottom row)]. For each type, sample flags are displayed in ascending
order of $d_H$ along the corresponding axis.}
\label{figure2}
\end{figure}
Altering the nature and strength of inter-cellular interactions, we observe a diversity of resulting patterns
of distinct cell fates that differ from the flag obtained in the uncoupled case not only in terms of the
lengths of the individual chromatic regions, but also in terms of their number and sequential order.
To quantify the variation in the flags obtained from the different realizations, we characterize them
by (i) the number $n_{B}$ of fate boundaries,
which are defined by adjacent cells having different fates, and (ii) a Hamming distance $d_H$ to the
idealized flag (obtained in the absence of coupling), determined by enumerating the number of cells whose fates are different in the two flags.
Depending on whether NICD up or downregulates the expression of the patterning genes, we obtain two qualitatively different outcomes.
While repression of B, W, R almost always results in flags having two boundaries
[Fig.~\ref{figure2}~(a-b)],
promoting their expression yields a much wider range of $n_B$
[Fig.~\ref{figure2}~(c-d)].
Furthermore, the flags generated for coupling types I and II are typically closer (in terms of $d_H$) to the idealized flag
as compared to types III and IV.
\begin{figure}[htbp!]
\includegraphics[width=0.5\textwidth]{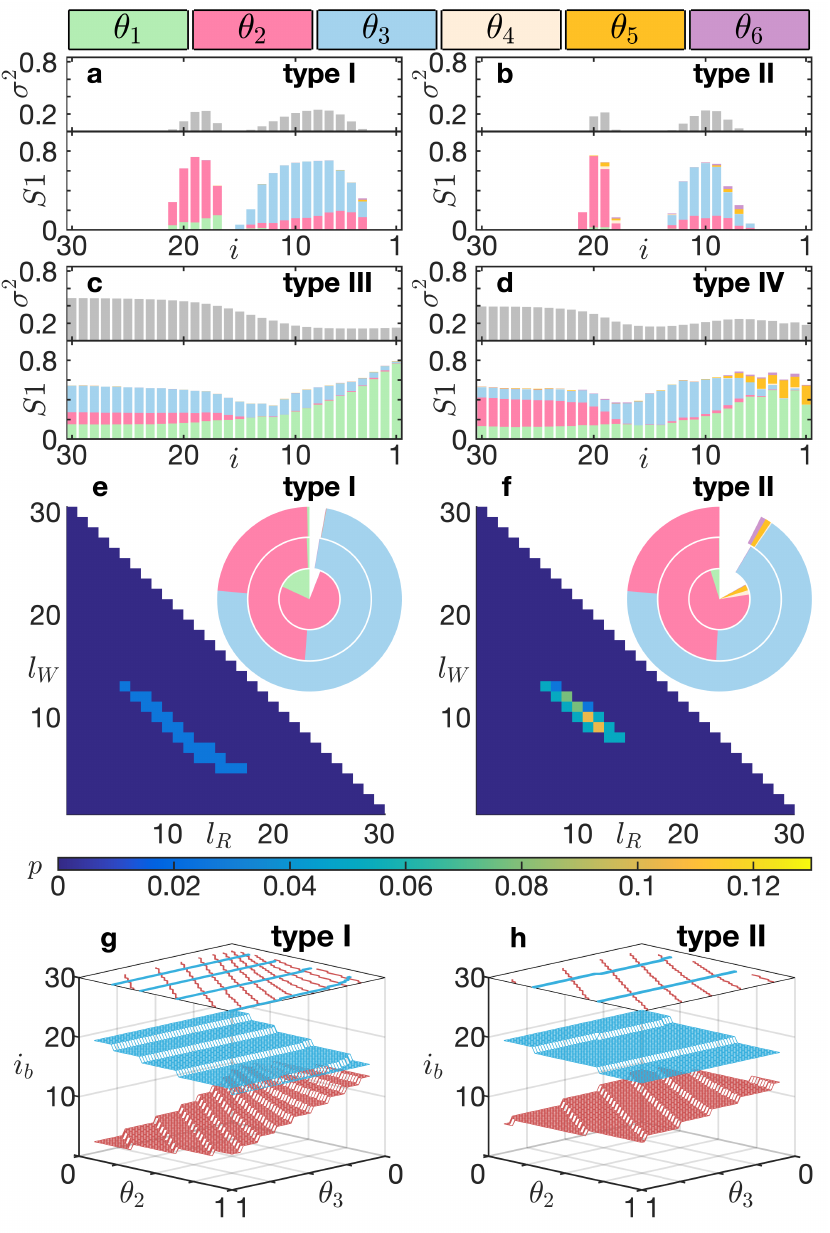}
\caption{\textbf{Sensitivity of the flags to inter-cellular coupling parameters.}
(a-d) Dependence of the variation in cell fates on the spatial location of each cell in a $1$D domain,
as well as, the differential contributions of the coupling parameters $\Theta$ to the variation, for the four types of
Notch-mediated interactions. The top half of each panel shows the variance $\sigma^2$ of the discrete variable
representing the three possible fates (blue/white/red) that a cell can attain. The bottom halves display the fraction of the variance that
can be accounted for by independently varying each of the parameters (colored according to the legend), as quantified by
the first order sensitivity index $S1$.
When NICD downregulates the patterning genes (a-b), most of the variation is localized around the 
two fate boundaries of the
uncoupled case and is sensitive to changes in $\theta_2$ and $\theta_3$.
In contrast, variation is seen across the domain when NICD upregulates the genes (c-d), with most of the contribution from
$\theta_1$, $\theta_2$ and $\theta_3$.
(e-f) Focusing on types I, II for which chromatic order and $n_B$ of the flags are invariant, we observe
that the lengths of the red and white segments ($l_R$ and $l_W$, respectively) are narrowly distributed around those
in the uncoupled case ($l^*_R = 10$, $l^*_W =10$).
The sensitivity of the segment lengths to the parameters $\Theta$ are shown in the concentric piecharts
(outer: $l_R$, middle: $l_W$, inner: $l_B$).
(g-h) The dependence of the location $i_b$ of the two fate boundaries (shown in blue and red, respectively)
on the parameters $\theta_2$ and $\theta_3$, with contour lines shown at the top.
}
\label{figure3}
\end{figure}

In contrast to the parameters governing the regulation of B, W and R by NICD, those
associated with modulating the effect of the patterning genes on ligand production
appear to have little or no effect on the resulting flags.
We use a variance-based sensitivity analysis technique to quantify the contribution of each
of these parameters in determining the cell fates~\cite{Saltelli1995}.
We consider the final state of each cell $i$ comprising the domain to be represented by a discrete scalar
variable $F_i \in \{0,1,2\}$ corresponding to blue, white and red.
Prior to quantifying the role played by the parameters $\Theta$ at each cell, we quantify
the variance ($\sigma^2$) in the fate $F_i$ across the different realizations
[Fig.~\ref{figure3}~(a-d), upper panels].
For coupling types I and II, we note that $\sigma^2$ is negligible throughout the array,
except around the location of the two fate boundaries in the idealized flag.
In contrast, $\sigma^2$ has a finite value at all locations in coupling types III and IV.
The contribution of the different parameters $\theta_{1,\ldots,6}$ to the observed variation in the fate
of each cell is measured by the respective first-order sensitivity indices $S1$,
expressed as the variance of $\langle F_i | \theta_j \rangle_{\theta_{k (\neq j)}}$
normalized by $\sigma^2$ (see SI).
Fig.~\ref{figure3}~(a-d, lower panels) show that only $\theta_2$ and $\theta_3$
contribute significantly in all coupling types, while for
coupling types III and IV, $\theta_1$ also plays an important role.
We note that the bulk of the variation in $F_i$ can be explained by $S1$ alone,
suggesting that the observed diversity can be largely explained in terms of
the independent actions of each parameter.

As flags that do not conserve $n_B$ or the chromatic order
of the idealized flag represent pronounced aberrations that are undesirable in the
context of development, we focus on coupling types I and II that are extremely
unlikely to generate such flags.
Indeed, the localization of variation in cell fates for these coupling types is consistent
with the resulting flags typically having low $d_H$ (see Fig.~\ref{figure2}).
Moreover, almost all of them have $n_B=2$, which allows the flags to be
uniquely specified by the lengths of any two out of the three chromatic regions.
Fig.~\ref{figure3}~(e-f) shows that the joint distribution of
the lengths $l_R$, $l_W$ of the regions having red and white fates, respectively,
is concentrated around that of the flag obtained in absence of coupling
(viz., $l_R=l_W=10$ for an array of 30 cells) for both coupling types.
The outer, middle and inner rings in the adjoining piecharts represent the contribution of each parameter $\theta_{1,\dots,6}$
to the variation observed in $l_R$, $l_W$ and $l_B$, respectively.
This is quantified by the corresponding first-order sensitivity indices, expressed in
terms of the angles subtended by each of the colored segments representing
the different parameters. Note that
the bulk of the observed variance in the lengths can be attributed to
changes in each of the parameters, independent of the others.
As $\theta_2$ and $\theta_3$ appear to be almost exclusively responsible
for the observed variation in the flags, in Fig.~\ref{figure3}~(g-h) we explicitly show how the locations
of the two boundaries $i_b$ between R, W (red) and W, B (blue) change on varying these
two parameters. For both coupling types,
increasing $\theta_2$ is observed to expand both the red and blue region at the
expense of the white region in the middle, while increasing $\theta_3$ results
in reduction of the red region but with little impact on the W-B boundary.
Thus, the variation in the flags resulting from
down-regulation by NICD of the genes forming the morphogen interpretation module
can be
explained by using only a pair of parameters controlling
the repression of W and R genes, respectively.
The predicted alterations in the resulting flag upon changing Notch expression can be tested experimentally to validate the role of inter-cellular interactions in determining the spatial pattern of cell fates outlined here.

To conclude, our work reveals that juxtacrine signaling between cells could play a key role in adaptively regulating cellular differentiation
that results in morphogenesis. While the diffusing morphogen, by setting up a global field
acts as the signal triggering the breaking of the intrinsic symmetry, and the
gene regulatory circuit forming the interpretation module translates the local morphogen
concentration into the eventual cell fates, inter-cellular interactions allow the information
from the environment of each cell to be incorporated into the process. Apart from
their utility in correcting for fluctuations in the signal in the presence of noise~\cite{Sprinzak2011}, such
an intermediate-scale process can increase the robustness of the system in
generating the desired flag by compensating for mutations affecting the production and/or
interpretation of the morphogen. We show this by using a modeling approach that
integrates two apparently
disparate paradigms for investigating biological pattern formation~\cite{Lander2011}, namely, that of
boundary-organized mechanisms involving
a pre-pattern such as a morphogen concentration gradient,
and self-organized mechanisms involving interactions between constituents~\cite{Green2015}.
\begin{acknowledgments}
We would like to thank James P. Sethna for helpful discussions.
SNM has been supported by the IMSc Complex Systems Project (12th
Plan), and the Center of Excellence in Complex Systems and Data
Science, both funded by the Department of Atomic Energy, Government of
India. The simulations required for this work were
supported by IMSc High Performance Computing facility (hpc.imsc.res.in) [Nandadevi].
\end{acknowledgments}

\clearpage
\onecolumngrid

\setcounter{figure}{0}
\renewcommand\thefigure{S\arabic{figure}}
\renewcommand\thetable{S\arabic{table}}

\vspace{1cm}

\begin{center}
\textbf{\large{SUPPLEMENTARY INFORMATION}}\\

\vspace{0.5cm}
\textbf{\large{Contact-mediated cellular communication 
supplements positional information to regulate spatial patterning during development}}

\vspace{0.5cm}
\textbf{Chandrashekar Kuyyamudi, Shakti N. Menon and Sitabhra Sinha}
\end{center}
\section*{List of Supplementary Figures}

\vspace{-0.25cm}
\begin{enumerate}
\item Fig S1: The final expression levels $B$, $W$ and $R$ of the patterning genes, as well as,
the concentrations of the Notch ligand, $L$, and NICD, $N^b$, for each of
the coupling types..
\item Fig S2: Spatio-temporal evolution of the expression levels of the patterning genes
B, W, and R at different cells in a 1D array comprising $30$ cells.
\item Fig S3: The effect of differential expression of Notch on cell fates.
\item Fig S4: ``Sloppy parameter sensitivity'' of the flags to inter-cellular coupling.
\end{enumerate}

\vspace{-0.2cm}
\section*{Model parameter values}

\vspace{-0.5cm}
\begin{table}[h]
\centering
\begin{tabular}{|c|c|c|c|c|c|c|c|c|c|c|c|c|c|c|c|c|c|c|c|}\hline
parameter& $S_M(0)$&$\lambda_M$ &$\alpha$&$\beta$&$\gamma$& $\beta_L$&$\beta_{N^b}$& $K$ & $K_N$ & $k_1$ & $k_2$ & $k_3$ & $\tau_L$ & $\tau_{N^b}$ &$h_1$&$h_2$&$h_3$&$h_4$&$h_5$\\\hline
value&100.0&0.3&4.0&6.3&5.0&5.0&5.0&1.0&1.0&1.0&1.0&1.0&1.0&1.0&6&2&5&1&1\\\hline
\end{tabular}
\caption{The values for the model parameters used for all simulation results reported (unless specified otherwise).}
\label{tableS1}
\end{table}

\vspace{-0.5cm}
\section*{Temporal evolution and parameter dependence of the patterns
representing the sequential arrangement of distinct cell fates}
\begin{figure}[H]
\includegraphics[width = \textwidth]{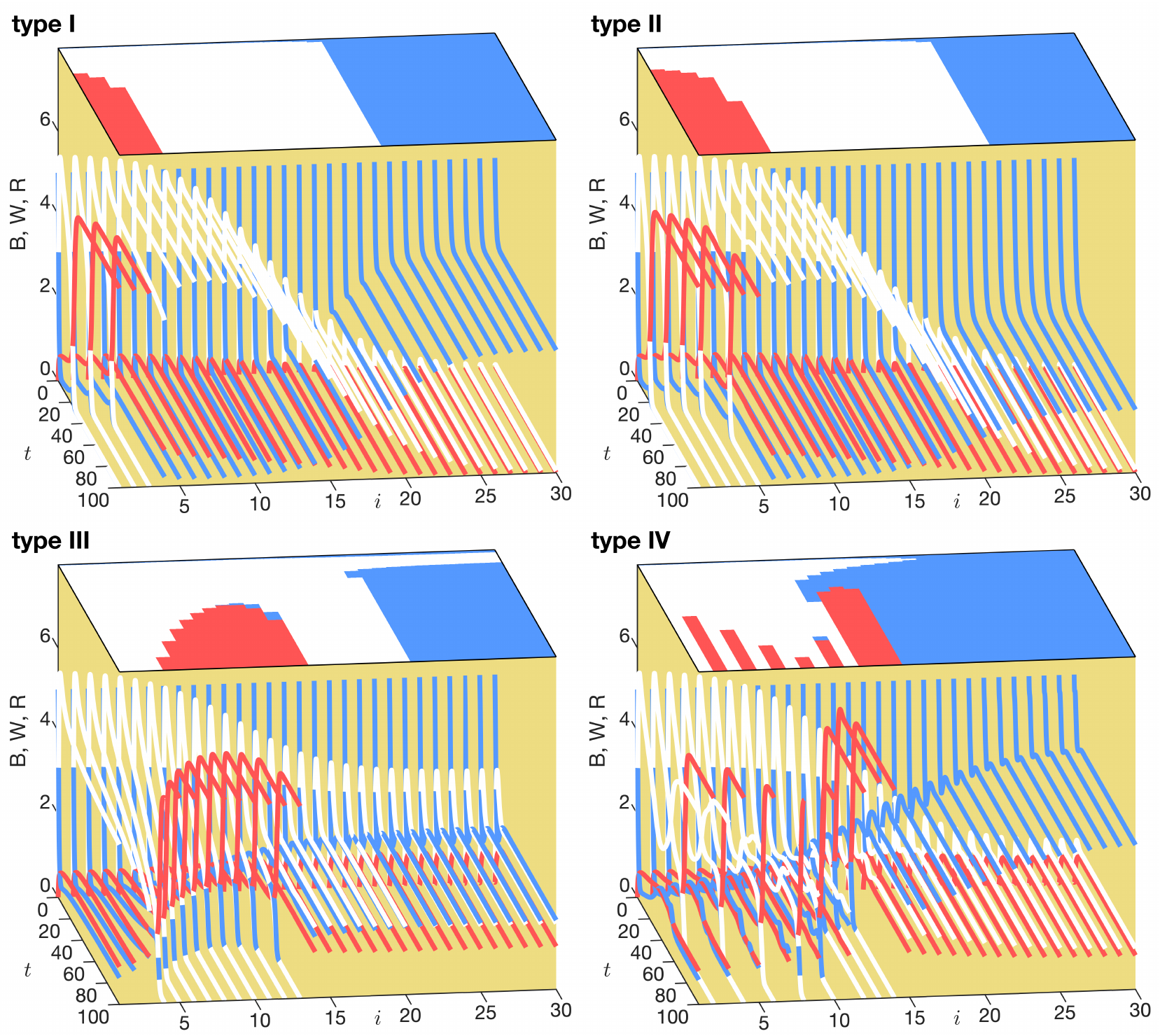}
\caption{{\bf Spatio-temporal evolution of the expression levels of the patterning genes
B, W, and R at different cells in a 1D array comprising $30$ cells.}
The resulting cell fates are indicated on the top surface of each panel corresponding to
the coupling types I-IV.
For each coupling type,
a representative parameter set $\Theta$ is used for the simulation.
Note that, all cells initially exhibit high expression levels for B.}
\label{figs1}
\end{figure}

The fate of each cell in the linear array we consider in our simulation is determined by
the expression levels of the three patterning genes B,W and R at $t_{max}$ = 100.
Gene B is chosen to be the pre-patterning gene, such that its expression level is high ($=5$
arb. units) initially while the initial expression level of the genes W and R is $0$.
Fig.~\ref{figs1} shows for each of the coupling types considered in the main text,
the time evolution of the
expression levels for B, W and R with a representative set of values chosen for the
coupling parameters. For coupling types I and II, B continues to be
the maximally expressed gene in cells that have relatively low exposure to the morphogen,
while in cells subject to intermediate and high morphogen concentrations W and R genes
(respectively) are the maximally expressed ones. In case of types III and IV,
more than two fate boundaries can emerge and the chromatic order seen in the uncoupled
system is not conserved.
This implies that the coupling types III and IV give rise to a large diversity of flags,
compared to types I and II.

\begin{figure}[H]
\includegraphics[width = \textwidth]{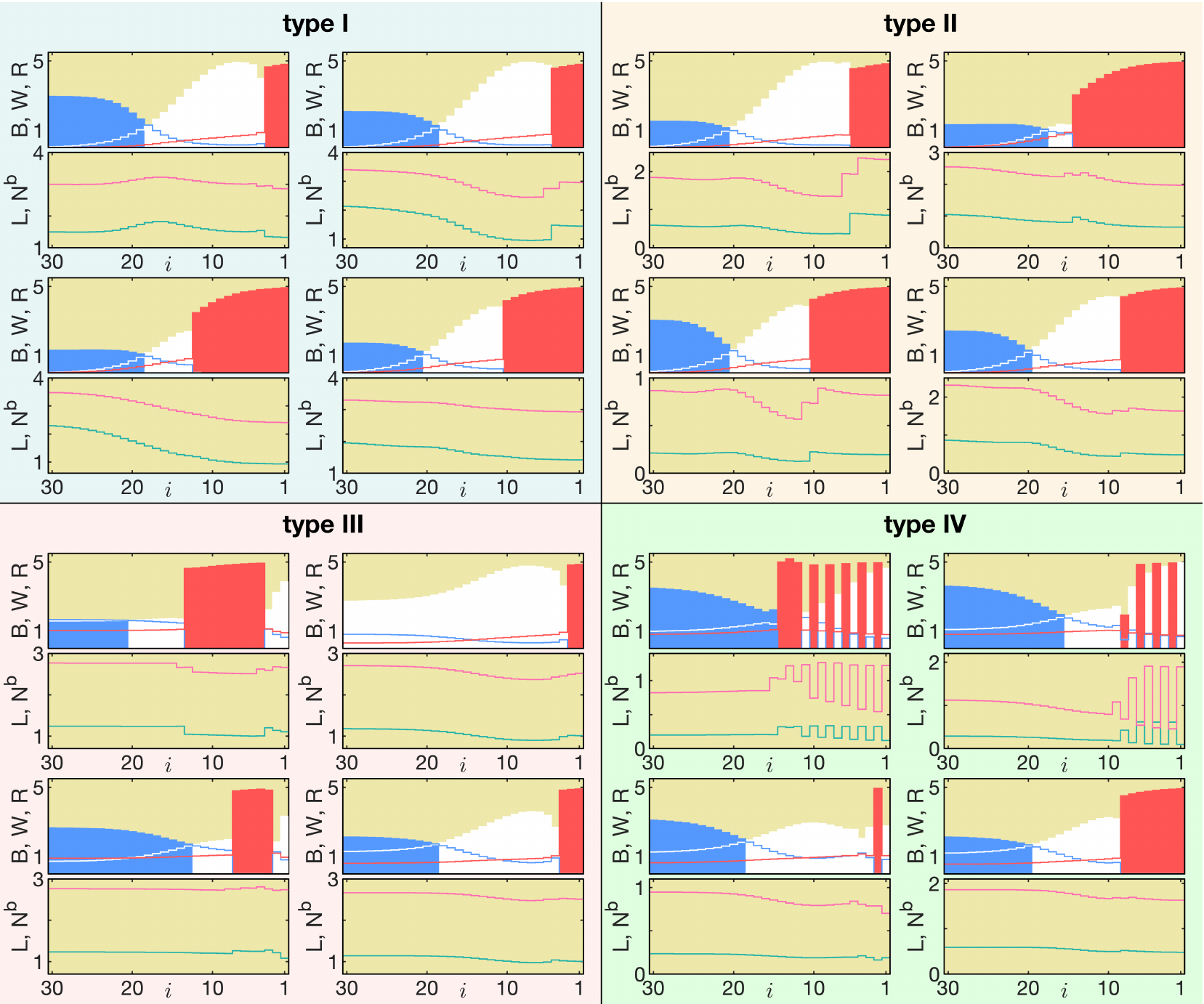}
\caption{{\bf The final expression levels $B$, $W$ and $R$ of the patterning genes, as well as,
the concentrations of the Notch ligand, $L$, and NICD, $N^b$, for each of
the coupling types.}
The maximally expressed gene at each cell in the 1D array determines its fate,
indicated by the colors blue, white or red.
For each coupling type, results obtained using different choices of values for the parameter
set $\Theta$ are shown.}
\label{figs2}
\end{figure}

Fig.~\ref{figs2} shows the final expression levels of the three pattering genes B, W, R in
all the cells of the linear array, along with the concentrations of the
Notch ligand, $L$, and the Notch intra-cellular domain NICD, $N^b$, for
four representative sets of parameter values for each of the four
coupling types.

\begin{figure}[H]
\includegraphics[width = \textwidth]{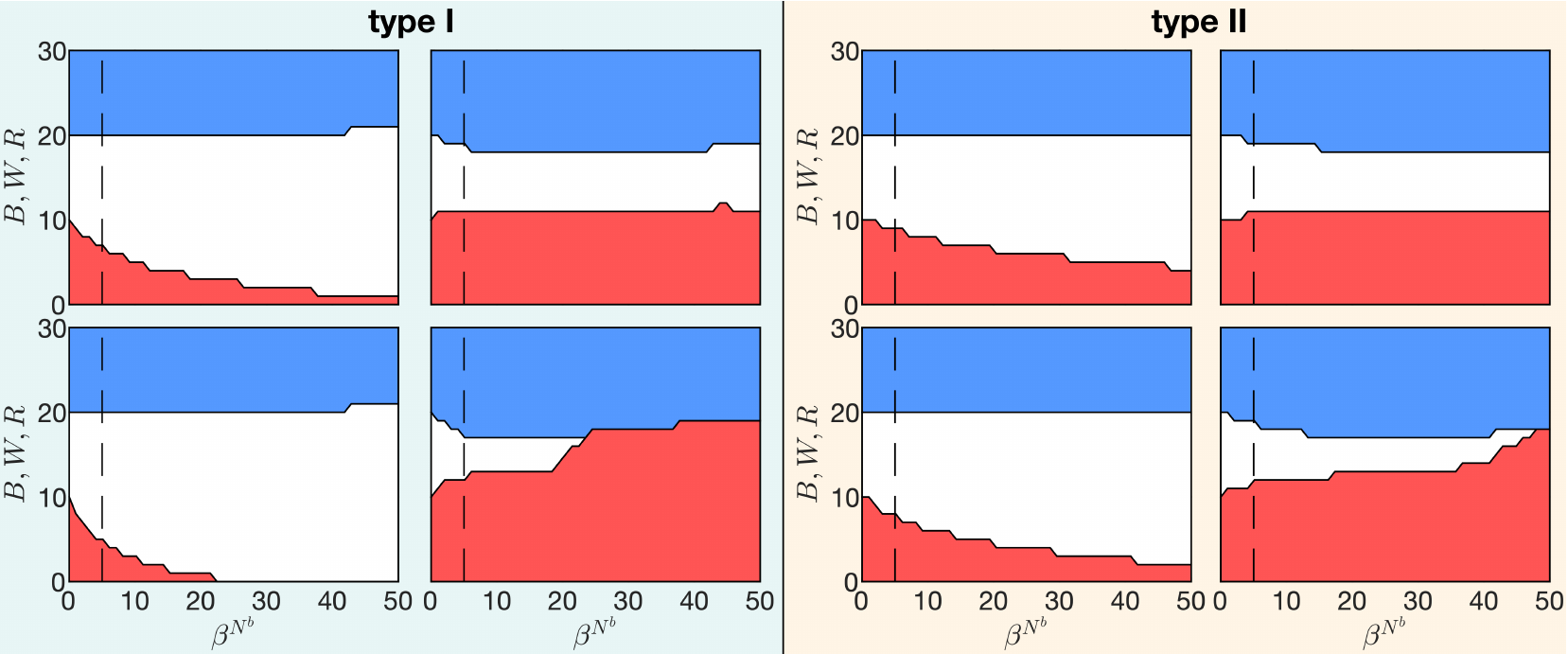}
\caption{\textbf{The effect of differential expression of Notch on cell fates.}
The variation of the spatial extent of the three chromatic regions (indicated using the colors
blue, white and red) with the NICD maximal production rate, $\beta^{N^b}$, for coupling
types I and II which preserve the chromatic order and number of boundaries seen
in the uncoupled system. For all results shown in the main text, we have chosen
$\beta^{N^b} = 5$ (indicated using a broken line). The result of under or over-expression of Notch relative to this value
is shown for different choices of values for the parameter set $\Theta$ in the case
of each coupling type.}
\label{figs3}
\end{figure}

We have considered the maximal production rate of NICD, $\beta^{N^b}$ to be $5$ for
all the results described in the main text.
To understand the implications of the over or under-expression of Notch for the results of
our model, we have also performed simulations by considering a
a wide range of values of $\beta^{N^b}$. We consider only coupling types I and II which
conserve the number of boundaries and the chromatic order observed in the uncoupled
system, ensuring that pathological patterns of cell fates do not arise for $\beta^{N^b}=5$.
Fig.~\ref{figs3} shows the variation in length of the chromatic regions as a
function of $\beta^{N^b}$ for four different sets of coupling parameter values for
each of the two coupling types. It is evident that, depending on the values of the
coupling parameters chosen, the higher expression of Notch can lead to either increase or
reduction of the red region, while the width of the blue region remains largely invariant.
This result suggests that experimental observation of the effect of under or over-expression
of Notch on the length of the regions with different cell fates can provide us
with information about the type of coupling and the strength of interactions in
different biological systems.

\section*{Parameter Sensitivity}
\subsection*{Variance-based sensitivity analysis}
To quantify the contribution of the parameters governing the regulation of
the patterning genes by NICD in determining the cell fates, we have used a variance-based
sensitivity analysis often referred to as the Sobol method~\cite{Saltelli1995}.
This method can be used to investigate the effect of varying any one of the parameters,
or a pair of them at a time, or any other higher-order combinations.
We consider a system that yields
a scalar output as a function of parameters $\theta_j$,
\begin{equation}
Y = f(\theta_1,\theta_2,..,\theta_k)\,.
\end{equation}
The first-order sensitivity index, which corresponds to the fraction of
the total variation in $Y$ that can be attributed to varying only
$\theta_j$, keeping the other parameters fixed, is defined as
\begin{equation}
S1 (j) = \frac{V_{\theta_i}(E_{\theta_{-j}}(Y|\theta_j))}{V(Y)}\,.
\end{equation}
Here, $-j$ refers to all other parameters except $j$. For each cell this sensitivity
index is, by definition, bound within the range $0\leq S1 (j) \leq 1$.
Note that the fraction of the total variance in the output variable that can be
explained by changing one parameter at a time is given by the sum over
all the first-order sensitivity indices,
$\sum_{j}{S1 (j)}\leq 1$.



As mentioned in the main text, we quantitatively investigate
the role of the parameter set $\Theta$ in determining the
final state of the cells $i$ ($i=1,\dots,30$) represented by a discrete scalar variable,
$F_i \in \{0,1,2\}$. The contribution of each of $\theta_k$ ($k=1,\ldots,6$)
to the observed variation in cell fates is measured by the respective
first-order sensitivity indices $S1$, expressed as the the variance of $\langle F_i | \theta_j \rangle_{\theta_{k (\neq j)}}$ normalized by $\sigma^2$ ($i= 1,\ldots,30$).
In case of the coupling types~I
and II, we almost always observe two boundaries separating the distinct
regions. This allows us to uniquely identify a resultant flag by specifying
$l_R$, $l_W$ and $l_B$, which correspond to the lengths of the red, white
and blue regions, respectively, and which can take discrete values in the range
$[0,30]$. Corresponding to these three scalar variables, we obtain first-order
sensitivity indices $S1^{R}$, $S1^{W}$ and $S1^{B}$, using quasi Monte Carlo
methods.

\begin{figure}[H]
\includegraphics[width = \textwidth]{Figures/FigS6.pdf}
\caption{\textbf{``Sloppy parameter sensitivity'' of the flags to inter-cellular coupling.}
The concentric pie charts (a: type I, b: type II) show the fractional contribution of the
components of the parameter set $\Theta$, represented by colors indicated in the key at the
top, to the sensitivity of the system to
parameter variation. This is expressed in terms of the spectral characteristics of the
Hessian matrix, whose eigenvalues $\lambda$ obtained for a specific parameter set for each
coupling type are shown at
the bottom.
For both coupling types, the two largest eigenvalues are comparable in magnitude
while there is a large gap between these and the subsequent eigenvalues.
The bar plots (c: type I, d: type II) show the probability
distribution [in logarithmic scale] of the exponential function of entropy $S$ for the
eigenvector components of the Hessian calculated for an ensemble of $10^3$ randomly chosen
parameter sets $\Theta$. Note that, as exp($S$) provides a measure for the number of dominant components in an eigenvector, most eigenvectors are dominated by a single
component. This is in agreement with the observation that almost all of the concentric shells in
the pie charts show the predominance of one color.
The aggregated contribution $w$ of each parameter to the eigenvectors (e: type I,
f: type II) indicates that, consistent with the variance-based sensitivity analysis reported
in the main text, $\theta_2$ and $\theta_3$ are almost exclusively responsible
for the observed variation in the flags.}
\label{figs4}
\end{figure}

\subsection*{``Sloppy parameter sensitivity'' analysis}
As a supplement to the variance-based sensitivity analysis, we have characterized
the sensitivity of the model output to variation of the parameters $\Theta = \{\theta_1, \ldots,
\theta_6\}$
using ``sloppy model
analysis''~\cite{Brown2003,Waterfall2006,Gutenkunst2007}.

This is done by varying each parameter over a relevant range and
calculating a Jacobian matrix that captures the variation of
output variables of interest. The Jacobian is then used to
obtain a Hessian matrix whose spectrum indicates the
sensitivity of the system to each of the parameters.\\

The output variables that we use to characterize the
sloppiness of the model are the lengths of the blue, white
and red regions. These are specified using the 3-tuple $(B,W,R)$, each of which can take
integer values between $0$ and $30$, subject to the constraint that $B+W+R=30$.
In conventional sloppy analysis of models, a specific set of parameter values is chosen as the
reference set $\Theta^{*}$ and the results of variations from this set are then investigated.
As in our model, there is no such privileged parameter set, we have carried out the analysis
using several different $\Theta^{*}$ obtained by randomly choosing
values of each of the parameters from their respective ranges.\\

To quantify the sloppiness of our model system,
we compare the lengths of the regions $(B^{*},W^{*},R^{*})$
obtained using a given set $\Theta^{*}$ with those obtained
using perturbed parameter sets. Each perturbed set
$\Theta_{i,j}$ is obtained by independently varying the
value of the parameter $\theta_j$ in the set $\Theta^{*}$
over the relevant range in small steps of $\Delta\theta$,
while keeping values of the other five parameters fixed.
Thus, for a given choice of $\Theta^{*}$ and $\Theta_{i,j}$, the residue is
\begin{equation}
\mathcal{R}_{i,j} = \sqrt{(B_{i,j}-B^*)^2 + (W_{i,j}-W^*)^2 + (R_{i,j}-R^*)^2}\,,
\end{equation}
where $(B_{i,j},W_{i,j},R_{i,j})$ denotes the lengths of the regions
obtained using the parameter set $\Theta_{i,j}$.
This is then used to obtain the Jacobian matrix {\bf J}
as
\begin{equation}
J_{i,j} = \frac{\mathcal{R}_{i+1,j} - \mathcal{R}_{i,j}}{\Delta \theta}\,,
\end{equation}
which comprises $(k-1)$ rows and $6$ columns, where $k$ is the number of perturbed parameter
sets considered. This subsequently
yields the $6 \times 6$ Hessian matrix
\begin{equation}
\mathcal{H} = J^{T}J.
\end{equation}
We calculate the eigen spectrum of
$\mathcal{H}$ and normalize the eigenvalues by the largest eigenvalue.
The number of ``sloppy'' directions corresponds to the
number of eigenvalues whose normalized magnitude $\ll 1$.
The eigenvector corresponding to an eigenvalue whose magnitude is very small
represents an axis in the $6$-dimensional
parameter space along which any variation has relatively little impact on the output.
Our results indicate that most eigenvectors have a single large component.
\\

We obtain the Hessian spectra for each of $10^3$ different
random choices of $\Theta^*$ (Fig.~\ref{figs4}).
Using these, we establish a hierarchy of the
six parameters $\theta_j$ in terms of their ``weights'' defined as
\begin{equation}
w_{\theta_j} = \frac{1}{N_{\Theta^*}} \sum_{\Theta^*}\sum_{k=1}^{6}{\lambda^k v^{k}_{\theta_j} * v^{k}_{\theta_j}}\,,
\end{equation}
where the first summation is over all $10^3$ parameter
sets $\Theta^*$ and the second is over
the six components of the spectra of $\mathcal{H}$.
Further, $\lambda^k$ and $v^{k}_{\theta_j}$ are the
eigenvalue and eigenvector of the $k$-th component
of the Hessian spectra obtained upon varying the
parameter $\theta_j$.

\begin{thebibliography}{99}
\bibitem{note1}
The feedback loop arises from
the Notch intra-cellular domain (NICD) of a cell $i$ controlling the production
of NICD in the neighboring cell $j$ by affecting the production
of the corresponding ligand in cell $i$, which in turn results in NICD
of $j$ controlling that of $i$, thereby completing the loop.
\bibitem{note2}
We assume that the concentration of morphogens across the domain is implicitly set by a spatially uniform linear
synthesis-diffusion-degradation (SDD) model, which yields an exponential gradient in the morphogen concentration $M$ at its steady state.
We assume that $S_M$ is a proxy for the concentration of the signaling molecule that is triggered when morphogen
molecules bind successfully with the cell membrane receptors.
\bibitem{note3}
We assume the number of receptors in each cell to be sufficiently high such that
saturation is not reached.
\end{thebibliography}

\end{document}